\documentclass[twocolumn]{aastex631}
\usepackage{epsfig}
\usepackage{graphicx}
\usepackage{mathtools}
\usepackage{hyperref}
\usepackage{color}
\usepackage{amsmath}
\providecommand{\sorthelp}[1]{}
\usepackage{verbatim}
\usepackage{lipsum}
\usepackage{mwe}

\bibpunct{(}{)}{;}{a}{}{,}

\shorttitle{California vs. Orion\,A in 3D}
\shortauthors{Rezaei Kh. \& Kainulainen}

\begin{document}

\title{
3D shape explains star formation mystery of California and Orion\,A
}

\author[0000-0002-0612-6152]{Sara Rezaei Kh.}
\affiliation{Max Planck Institute for Astronomy (MPIA),  K\"onigstuhl 17, 69117 Heidelberg, Germany}
\affiliation{Chalmers University of Technology, Department of Space, Earth and Environment, 412 93 Gothenburg, Sweden}

\author[0000-0001-7764-3109]{Jouni Kainulainen}
\affiliation{Chalmers University of Technology, Department of Space, Earth and Environment, 412 93 Gothenburg, Sweden}

\email{sara@mpia.de}

\begin{abstract}
The new Gaia data release (EDR3) with improved astrometry has opened a new era in studying our Milky Way in fine detail. We use Gaia EDR3 astrometry together with 2MASS and WISE photometry to study two of the most massive molecular clouds in the solar vicinity: Orion\,A and California. Despite having remarkable similarities in the plane of the sky in terms of shape, size, and extinction, California has an order of magnitude lower star formation efficiency. We use our state-of-the-art dust mapping technique to derive the detailed three-dimensional (3D) structure of the two clouds, taking into account both distance and extinction uncertainties, and a full 3D spatial correlation between neighbouring points. We discover that, despite the apparent filamentary structure in the plane of the sky, California is a flat 120-pc-long sheet extending from 410 to 530 $pc$. We show that not only Orion\,A and California differ substantially in their 3D shapes, but also Orion\,A has considerably higher density substructures in 3D than California. This result presents a compelling reason why the two clouds have different star formation activities. We also demonstrate how the viewing angle of California can substantially change the cloud's position in the Kennicutt-Schmidt relation. This underlines the importance of 3D information in interpreting star formation relations and challenges studies that rely solely on the column density thresholds to determine star formation activities in molecular clouds. Finally, we provide accurate distance estimates to multiple lines of sight towards various parts of the two clouds.
\end{abstract}

\keywords{
ISM: structure --- ISM: clouds --- ISM: bubbles --- Molecular cloud: California, Orion\,A --- ISM: dust, extinction --- Star formation
}

\section{Introduction}\label{sec:intro}
Plane-of-the-sky studies have provided valuable information about the star-forming interstellar medium (ISM). However, our knowledge of the true, three-dimensional (3D) structure of molecular clouds remains limited, which hampers our understanding of how the clouds evolve and form stars. But now, thanks to the unprecedented astrometry from the Gaia mission \citep{Gaia16}, studying the 3D structure of molecular clouds has become possible \citep[e.g.,][]{Rezaei_Kh_20,Zucker_20,Leike_20,Grossscheld18,Kain21}. In this paper, we compare the 3D shapes of two molecular clouds and demonstrate how that knowledge has a fundamental effect in understanding their star formation activities.

The California and Orion\,A molecular clouds are two of the most massive giant molecular clouds (GMCs) within 500 $pc$ from the Sun. The two clouds are located relatively near one another and show comparable kinematics and similar filamentary shapes and sizes in the plane of the sky \citep{Lada_09}. However, their star formation activities differ significantly, with Orion\,A having an order of magnitude higher star formation rate than California \citep{Lada_09}. The difference triggers an immediate question about what regulates star formation and what role the individual cloud’s properties like shape and size play in setting its star formation activity.

To deepen our understanding of the contrasting star formation rates in California and Orion\,A, we investigate their 3D shape by mapping the dust distributions towards the two clouds. The detailed 3D dust distribution towards Orion\,A was already presented in \cite{Rezaei_Kh_20} where we revealed new information on its 3D shape, discovered a foreground dust component, and demonstrated the extending tail of the cloud to further distances as suggested by \cite{Grossscheld18}. Here, we look into the 3D distribution of the dust towards California to map its 3D substructures that remain hidden in the plane-of-the-sky data.

While there are a couple of local dust maps of the Milky Way, our current work towards California and Orion\,A is a unique map of its kind. \cite{Leike_20} provides a 3D dust map of the solar neighbourhood using Gaia DR2 and metric Gaussian variational inference which is then used by \cite{Zucker_21} to define cloud boundaries. However, \cite{Leike_20} provides data for all lines of sight (l.o.s) only up to 370 pc and for some l.o.s close to 500 pc, depending on the coordinates. Their results beyond 370 $pc$ need to be treated with extreme caution, as it reaches the boundaries of the map which affects the density, shape and distance to the clouds \citep{Zucker_21}. Apart from the coverage limits of \cite{Leike_20}, our 3D dust mapping technique has the advantage of taking into account distance uncertainties; as a result, it can exploit a full dataset like Gaia without the necessity to cut on noisy data. Those stars are typically located in dusty regions which are important for mapping the dense parts of the molecular clouds. Moreover, owing to the analytical solution for our posterior calculation, approximations are minimal; therefore, our method provides accurate and reliable estimates \citep[see,][for more details]{Rezaei_Kh_17,Rezaei_Kh_18b,Rezaei_Kh_20}.

\section{Method and Data} \label{sec:method}

Our 3D dust mapping technique has been extensively explained in \cite{Rezaei_Kh_17} and \cite{Rezaei_Kh_18b}. In addition, new developments have been introduced in \cite{Rezaei_Kh_20}. Here we briefly summarise the main aspects of our method.

Our technique consists of a non-parametric method that uses 3D positions of stars together with their line-of-sight (l.o.s) attenuation as the input data and predicts dust densities for arbitrary points in the same 3D space. We divide the l.o.s towards stars into 1D cells in order to model the integrated attenuation to each star as sum of the dust densities along its l.o.s by connecting all 1D cells using a Gaussian Process (GP) prior. GP takes into account the neighbouring correlations using a correlation length, ${\lambda}$. Another hyper-parameter of the model is the variance, ${\theta,}$ which sets the amplitude of density variations. All hyper-parameters are fixed based on the input data as explained in \cite{Rezaei_Kh_17, Rezaei_Kh_18b}. The final resolution of the map is set by the typical separation between input stars. Apart from considering the 3D neighbouring correlation, our method has the advantage of taking into account both distance and extinction uncertainties towards individual stars which result in producing robust estimates of the distance and 3D shapes of the molecular clouds. Despite common artefacts in most of the 3D dust maps, our results are devoid of discontinuity and ``fingers-of-God'' artefact.

The early instalment of the third Gaia data release (Gaia EDR3) consists of the full astrometric solution for around 1.5 billion sources, and a significant advantage over Gaia DR2 parallaxes with 30 percent increase in precision \citep{Gaia21}. We use the 3D positions of stars from Gaia EDR3; only limiting our sample to parallax uncertainty of less than 100 percent. Since we are dealing with large parallax uncertainties, we use the geometric distance estimates from \cite{Bailer-Jones21} to map the full shape of the cloud. 

Similar to \cite{Rezaei_Kh_20}, we use the Two Micron All-Sky Survey \citep[2MASS;][]{Skrutskie06} and the Wide-Field Infrared Survey Explorer \citep[WISE;][]{Wright10} photometry to calculate extinctions to individual stars using the Rayleigh-Jeans Colour Excess \citep[RJCE,][]{Majewski11} method, which provides us with the extinction in $K_{S}$ band. The crossmatch between 2MASS and WISE catalogues with Gaia EDR3 sources are provided on Gaia Archive\footnote{https://gea.esac.esa.int/archive/.}.

After calculating extinctions, we select our final sample based on the position of stars on the de-reddened colour-magnitude diagram in order to remove the outliers \citep[see][]{Rezaei_Kh_18a,Rezaei_Kh_20}. For our current work towards the California cloud, $155^{\circ} < l < 170^{\circ}$ and $-14^{\circ} < b < -6^{\circ}$, we use around 160\,000 stars as our final input sample. The hyper-parameters of the method for the aforementioned data are: cell size $= 5 pc$, ${\lambda = 20 pc}$, and ${\theta} = 4 {\times} {10}^{-8}$ ${pc}^{-2}$. It is important to note that the choice of hyperparameters are not completely arbitrary. As discussed in detail in \cite{Rezaei_Kh_17, Rezaei_Kh_18b}, the hyperparameters are calculated according to input data. The final resolution of the map is not set by these parameters, but rather the input data (see section \ref{mock} for an example).

\section{California in 3D} \label{sec:map}

\begin{figure*}
\resizebox{\hsize}{!}{\includegraphics[clip=true]{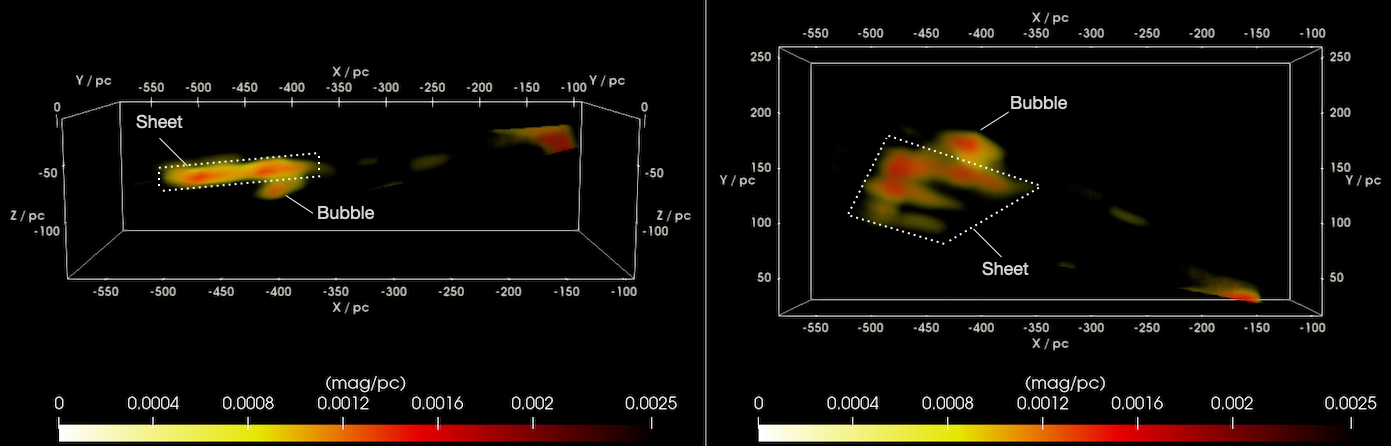}}
\caption{3D dust density predictions for California cloud projected from the Y (left panel) and the Z (righ panel) axis (perpendicular to the Galactic plane). More angles can be seen in the appendix (Fig. \ref{fig:3D_more}). The colour represents densities in $K_{s}$-band magnitude per parsec. The sun is at (0,0,0) and X increases towards the Galactic Centre. The dotted lines demonstrate what we refer to as a sheet. The predictions are made on regular grids for every 0.5 degrees in the Galactic $\emph{l}$ and $\emph{b}$, and every 5 $pc$ in distance. The 3D image is then produced by applying a smoothing kernel to handle the missing pixels. In order not to produce extra smoothing than that of the method, the length scale of the smoothing kernel is chosen to be much smaller (3 $pc$) than the correlation length. For illustration purposes, values below 0.0005 are set to be transparent. For $N_{H} ({cm}^{-2}) = 2 {\times} {10}^{22} A_{K}$, the dust density of 0.001 $mag/pc$ corresponds to gas volume density of $\sim 6$ $cm^{-3}$ (see Fig \ref{fig:los}). \label{fig:3d}}
\end{figure*}

\begin{figure*}
\resizebox{\hsize}{!}{\includegraphics[clip=true]{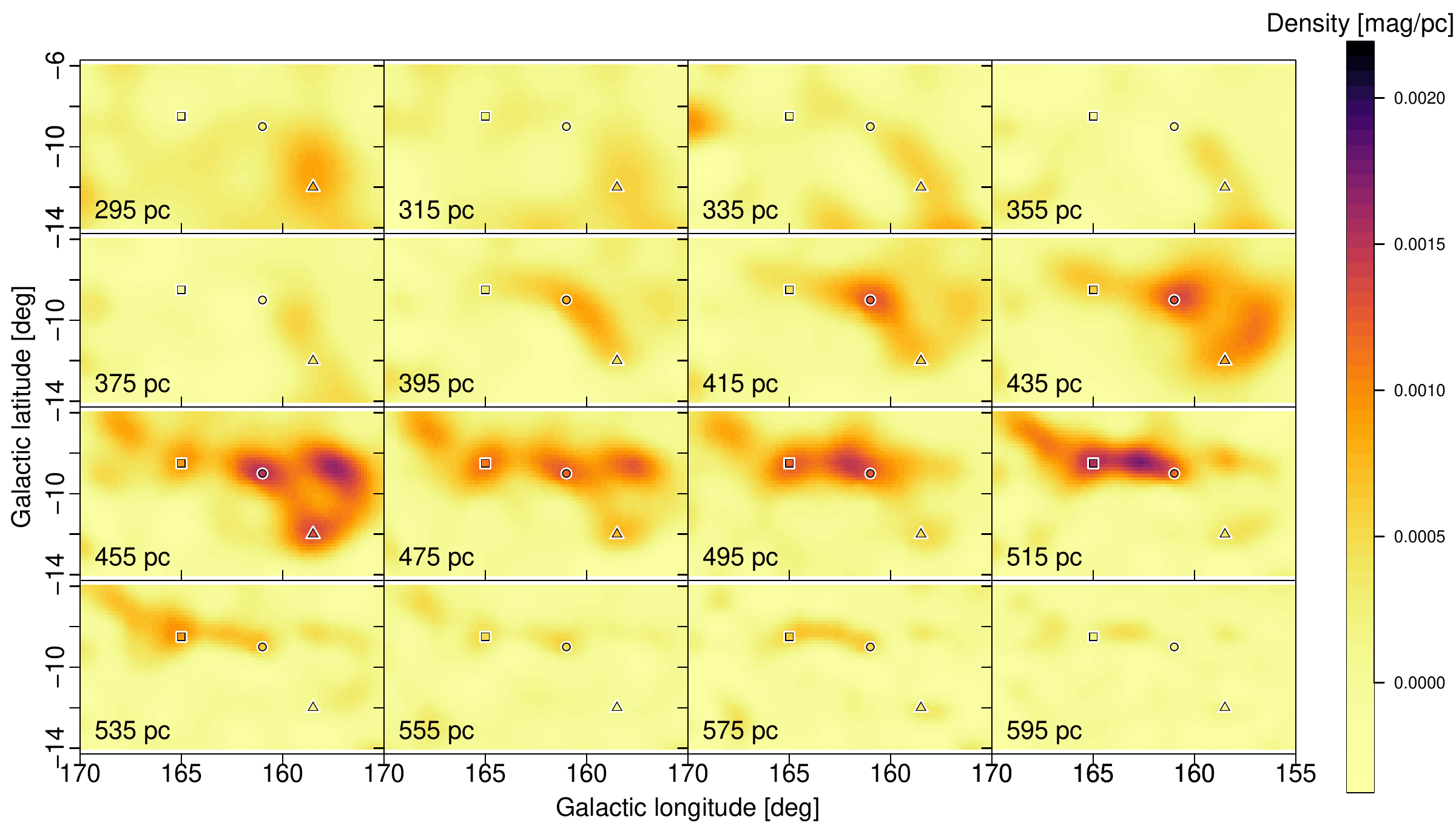}}
\caption{Dust density predictions in the plane of the sky. Each panel represents a slice through the cloud at fixed distances (every 20 $pc$). The bubble on the right side is apparent at 455-pc slice, while the rest of the cloud is seen through multiple panels demonstrating the extent of the sheet, especially towards $(l , b) = (161.5 , -8.5)$. The three symbols (square, circle, and triangle) represent specific l.o.s along the cloud investigated in Fig.~\ref{fig:los}. For illustration purposes, the image is smoothed with the scale length of 0.4 degrees. For $N_{H} ({cm}^{-2}) = 2 {\times} {10}^{22} A_{K}$, the dust density of 0.001 $mag/pc$ corresponds to gas volume density of $\sim 6$ $cm^{-3}$. \label{fig:slice}}
\end{figure*}

Figure \ref{fig:3d} shows the 3D distribution of dust towards the California region from two viewing angles. Rather than being a filament perpendicular to the l.o.s, California is an extended flat, sheet-like structure, with a bubble at one side, where higher-density structures form within the sheet. From the face-on view, the longest length of the sheet is approximately 120 $pc$, with the shortest length of about 80 pc, while from the edge-on view, the width of the sheet appears very narrow (as expected from the plane-of-the-sky observations). Figure \ref{fig:slice} demonstrates various components of the cloud at multiple distance slices: while the bubble centred at $\emph{l} = 158^{\circ} , \emph{b} = -10^{\circ}$ appears clearly at 455-pc panel and expands over only 25 $pc$ in radius, some other parts of the cloud extend along the sheet for over 100 $pc$.

\begin{figure}
\begin{center}
\includegraphics[width=0.50\textwidth, angle=0]{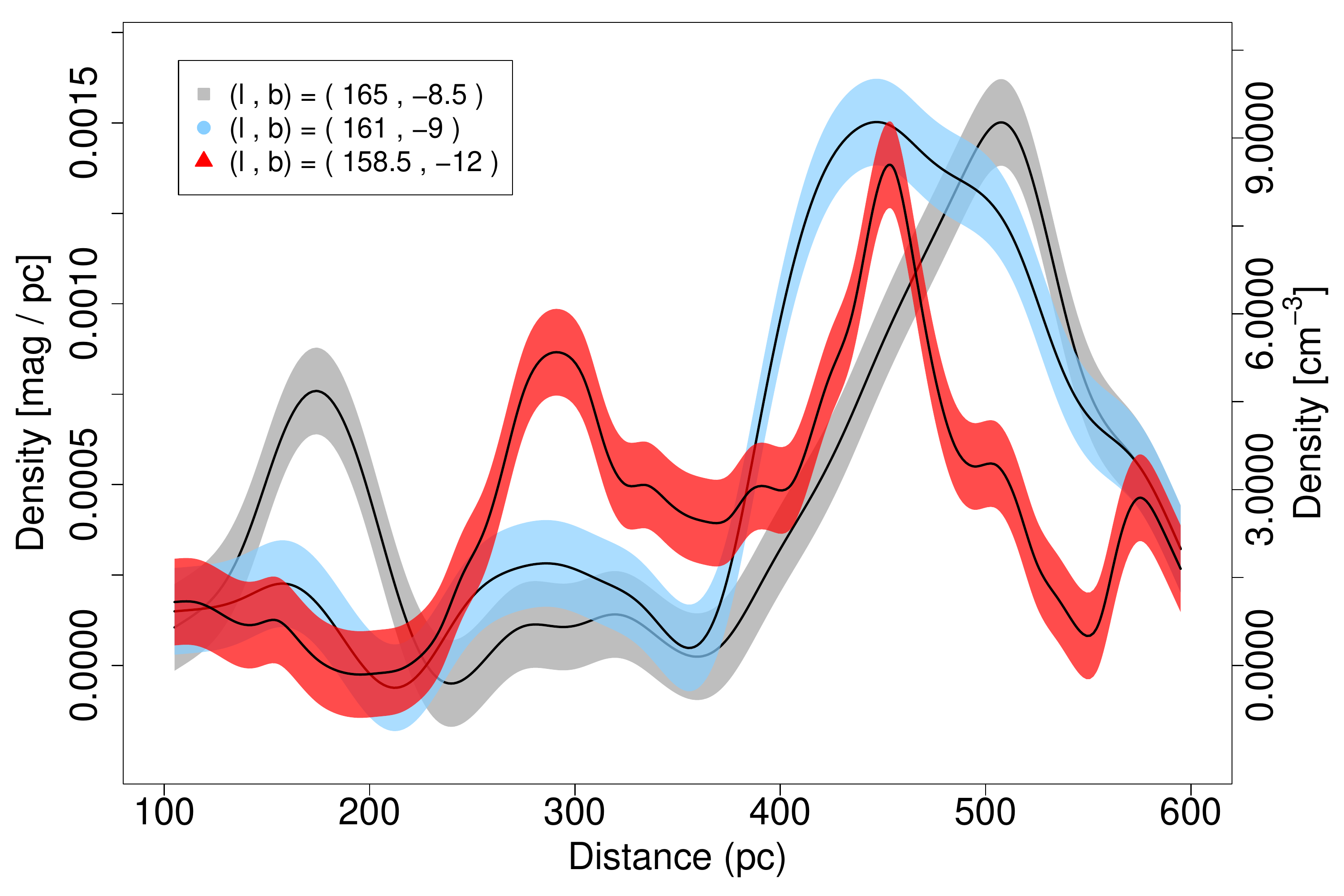}
\caption{Dust density vs. distance for three different l.o.s. towards California cloud (over-plotted on Fig.~\ref{fig:slice}). The black line shows the mean and the shades represent one standard deviation (also computed by the Gaussian process model). The red curve is towards the lowest part of the bubble, the blue curve represents the elongated length of the sheet, and the grey curve shows the density for a l.o.s towards the left (Eastern) end of the cloud. The density in the secondary axis is calculated assuming $N_{H} ({cm}^{-2}) = 2 {\times} {10}^{22} A_{K}$, which is an average value for the hydrogen column density in the literature \citep[e.g.][]{Reina73, Gorenstein75, Predehl95, Guver09}. \label{fig:los}}
\end{center}
\end{figure}

Having uncertainties on the dust density predictions is crucial for evaluating the significance and validity of different 3D sub-structures, as explained in \cite{Rezaei_Kh_20}. Figure \ref{fig:los} shows dust density predictions with their uncertainties as a function of distance for three l.o.s towards different parts of the cloud (also marked in Fig. \ref{fig:slice} in the plane of the sky). The red curve that is towards the lowest part of the bubble has a dominant, symmetrical over-density at 445 $pc$ and there is a shallower over-density in the foreground at 290 $pc$. The foreground over-density is in agreement with an increase in reddening in \cite{Green_19} map at the same distance, and an increase in density in \cite{Leike_20}. The blue and grey curves, however, recover a broader distribution that represents the extent of the sheet along the l.o.s. While the highest peak of dust density along $(l , b) = (161 , -9)$ (blue curve) is around 450 $pc$ \citep[as estimated by][]{Zucker_20}, it captures the most elongated part of the sheet extending from 410 $pc$ to 530 $pc$, giving it the length of about 120 $pc$. There is also a shallow over-density at closer distances towards this l.o.s which is related to the Taurus cloud on the foreground. As seen from the grey curve, $(l , b) = (165, -8.5)$, the peak density of the Eastern part of California is at a further distance of 514 $pc$.

It is important to note that the extension of the clouds along various l.o.s are driven by the input data; in fact, as explained in our previous works, our method is capable of capturing multi-scale density substructures that are smaller or larger than the input correlation length. In addition, the peak densities presented here are not representative of the high-density small-scale molecular gas but an average density within the resolution of the map; therefore much higher density cores can be located within our dust clumps. We demonstrate these points with a mock dataset in the Appendix \ref{mock}.

We can reconstruct the 2D extinction map towards California by integrating our predicted 3D dust densities along various l.o.s. As can be seen from Fig. \ref{fig:Ak}, the final 2D projected map reaches extinction of $\sim0.5$ magnitude in $K_{s}$ band ($\sim5$ $mag$ in $A_{V}$) and there is a clear resemblance to higher resolution maps (e.g. CO-based maps or NICEST/NICER extinction maps).

\section{California vs. Orion\,A} \label{sec:discussion}

\begin{figure*}
\resizebox{\hsize}{!}{\includegraphics[clip=true]{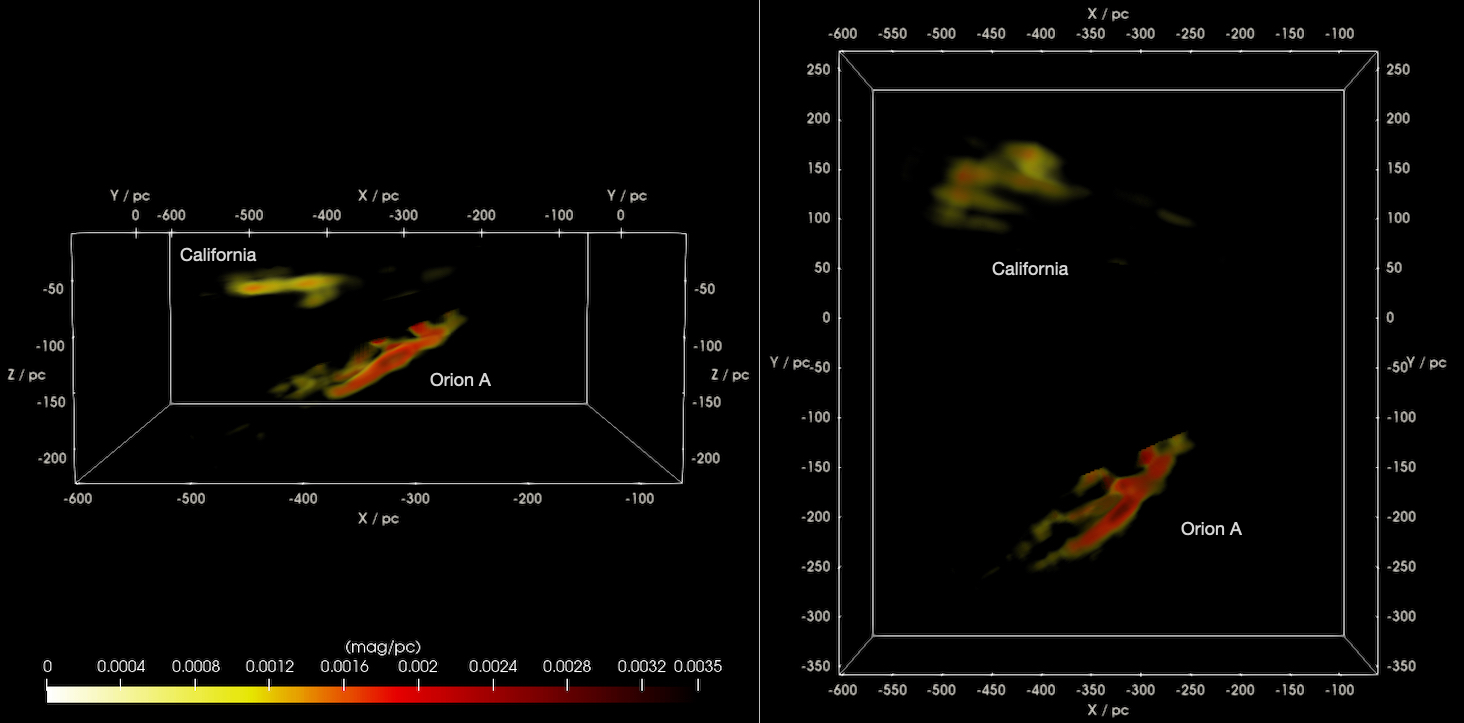}}
\caption{Two projected views of the 3D distribution of dust towards the California and Orion\,A clouds. The sun is at (0,0,0), the Z-axis is perpendicular to the Galactic plane, and X increases towards the Galactic Centre. For illustration purposes, values below 0.0005 are set to be transparent. \label{fig:both}}
\end{figure*}

Even though very similar in the plane of the sky \citep{Lada_09}, California's 3D structure appears very different to that of Orion\,A. Within our resolution limit, only one apparent feedback-driven substructure is seen in California (a bubble on the Western side), while Orion\,A has several such structures. Beyond this substructure, California is distributed along a flat sheet, while Orion\,A is more filamentary also in 3D. The two clouds are shown together in Fig. \ref{fig:both}. The substructures in Orion\,A are more complex than in California, likely caused by feedback processes and multiple episodes of star formation in the region \citep[e.g.][]{Rezaei_Kh_20, Schlafly15}. In addition, substructures in Orion\,A have higher densities than in California (see Fig. \ref{fig:both}). Since the shallow over-densities in California are spread along the flat sheet, they add up in the plane of the sky to represent extinction properties similar to those of Orion\,A. However, the 3D view reveals the dramatic differences between the clouds.

In addition, other studies have estimated the age of California to be around 1-3 Myrs \citep[e.g.,][]{Wolk_10,Covey_10,Imara_17}, while Orion\,A seems to be in an older evolutionary stage with older stellar populations \citep[5 - 10 Myrs; e.g.,][]{Bouy14,Zari19} associated with its discovered foreground bubble \citep{Rezaei_Kh_20}, which seems to have triggered the next generation of star formation in the main Orion\,A filament in the background.

\begin{figure}
\centering
\includegraphics[width=0.50\textwidth, angle=0]{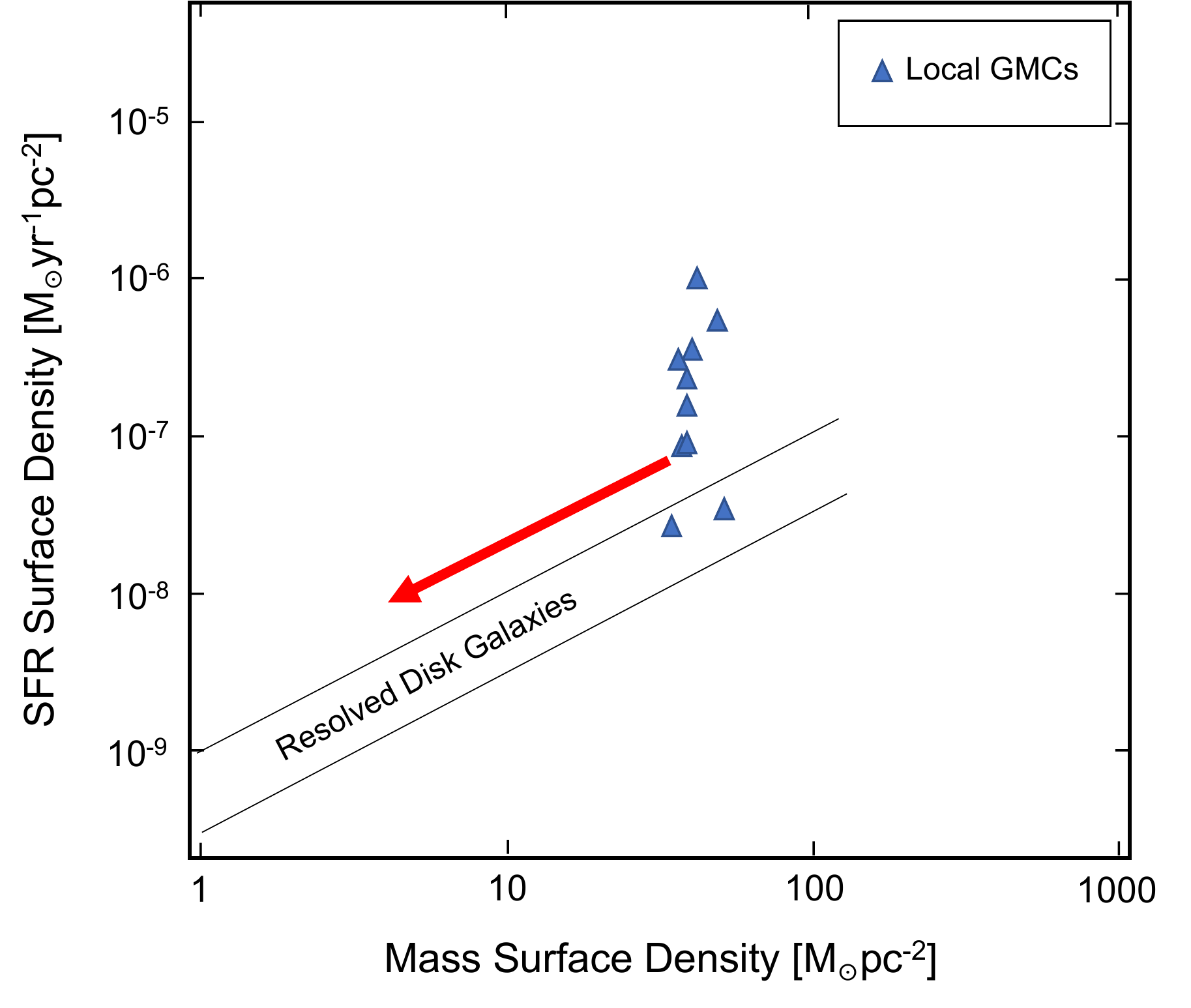}
\caption{Star formation surface density as a function of mass surface density (Kennicutt–Schmidt relation) for the local molecular clouds from \cite{Lada_13}. The red arrow shows how the location of a cloud like California would change on this diagram when seen from an angle perpendicular to the l.o.s. \label{fig:KS}}
\end{figure}

We also demonstrate that the knowledge of clouds’ 3D shape is an important factor in interpreting their star formation activities and in considering star formation relations such as the Kennicutt-Schmidt (KS) relation. Without observational biases, the mean surface density of the California cloud changes drastically depending on the viewing angle: the total mass of the cloud is the same irrespective of the viewing angle, but the surface area is an order of magnitude larger when viewed from the direction perpendicular to the Galactic plane (perpendicular to the cloud’s sheet-like morphology). The change in area is less dramatic for Orion\,A that is more filamentary than sheet-like. This effect has two potentially important consequences:

Firstly, the location of California in the KS-relation depends significantly on the viewing angle. The SFR in a molecular cloud in the Milky Way is calculated from the number of young stellar objects (YSOs) in the region, assuming an average mass and age for the YSOs. Since the number of observed YSOs doesn’t change with the angle, then SFR will remain constant. As the cloud area changes, but the mass and SFR remain constant, California can move in the relation along a line with the slope of unity (see Fig. \ref{fig:KS}).

How exactly the different structures of the clouds affect their location in the \emph{observed} KS-relation depends on the details of the observational techniques used to derive the masses and areas of the clouds. In \cite{Kain21}, we showed that especially the column density threshold used to define the clouds has a strong effect on this. Particularly, using plane-of-the-sky data, \cite{Lada_13} found that there is no KS-relation between the clouds in the Solar neighbourhood. Our result opens a door to speculate that the relation may well exist, but, depending on the size and morphology of the clouds, is not recovered from the plane-of-the-sky data. This could be either due to systematic effects in the cloud orientations, coincidence in the sampling of random orientation angles, or the details of the cloud definition process \citep[especially thresholding; ][]{Kain21}.

Secondly, our result also suggests a link between the 3D morphology of molecular clouds and their star formation activity. The plane-of-the-sky works indicate that the California and Orion\,A clouds are morphologically similar; the origin of the differing star formation activities has been attributed to differing amounts of dense gas \citep[e.g.,][]{Kainul_09,Lada_13,Kainul_17} and/or environmental effects \citep[e.g.,][]{Schlafly15}. Our work shows that the two clouds are \emph{not} morphologically similar, which immediately raises the question about the importance of morphology. Again, providing 3D shape information for a larger sample of clouds will be vital to establish this.

Finally, we provide the most up-to-date distance estimates to various l.o.s towards California and Orion\,A using Gaia EDR3 (table \ref{tab:dist}). We note that the geometric distance estimates from GEDR3 parallaxes by \cite{Bailer-Jones21} are over-estimated because their prior's scale length is larger than our targeted cloud distance closer than 500 $pc$. Therefore, after we map the full 3D shape of the cloud using all stars that match our input criteria (see section \ref{sec:method}), in order to get accurate distances to individual dust components along various l.o.s, we only use stars from Gaia EDR3 with parallax estimates better than 20 percent and directly invert them to get their distances.

There is a perfect agreement between our distance estimates and distance estimates of \cite{Zucker_20} for l.o.s towards where there is a prominent peak. However, where there are multiple peaks near each other along a l.o.s (like the foreground cloud in Orion\,A) or where there are no dominant peaks (e.g. elongations in California), our distance estimates differ. Although distances in \cite{Zucker_20} are predicted on higher spatial resolution than that of our work, it is possible that due to the nature of \cite{Zucker_20} technique, multiple close-by peaks and the physical shape of the clouds remain undetected in their work.
It is important to note that, as explained in \cite{Rezaei_Kh_20}, our predictions are made for fixed points in 3D; therefore, there is no direct uncertainty associated with our distances. However, our method provides the full probability distribution function for density predictions at each point which can then be propagated into distance uncertainties. Having done so, the typical uncertainty in our predicted distances are 10 $pc$.

\section{Concluding remarks} \label{sec:conclusion}

We have determined the 3D structure of the California molecular cloud using our advanced mapping technique that takes into account distance and extinction uncertainties to individual stars and considers correlations between neighbouring points in 3D. Having the full probability density function for each predicted point, i.e. mean and standard deviation of the predicted density, allows us to verify the results; therefore, our final products are robust 3D dust maps of the molecular clouds.

We demonstrated that California is a sheet-like structure extended around 120 $pc$ along the l.o.s. It consists of low-density substructures that align with the l.o.s resulting in higher column density in the plane of the sky, comparable to that of Orion\,A. The surface density of California, however, is much lower than Orion\,A from the viewing angle perpendicular to the Galactic plane. In addition, the substructures of the two clouds show substantial differences with Orion\,A being a filamentary structure with several feedback bubbles while California has only one isolated bubble on one side and largely resembles an unperturbed sheet. The two clouds, even though similar in the plane of the sky, are remarkably different in 3D. The dramatic differences likely contribute to the different star formation efficiencies of the clouds. Our results demonstrate the importance of 3D information in understanding the star formation activities of molecular clouds.

We also showed that the 3D information plays a dramatic role in setting the mean surface density of the clouds, which can then significantly affect their appearance in scaling relations such as their location in the KS relation. This further indicates that the column density thresholds are not necessarily reliable tools to determine core star formation activities in molecular clouds. We have presented the results for California in this paper, but plan to expand this study to more clouds in order to have a better understanding of star formation in the solar neighbourhood and its connection to extragalactic studies.

\begin{table*}
\caption{Distance estimates to Orion\,A and California using Gaia EDR3} 
\label{tab:dist}     
\centering                          
\begin{tabular}{c c c c c}  
\hline\hline                 
Name	& $l$	  [$^{\circ}$]	& $b$ [$^{\circ}$] 	&	Distance [$pc$] to peak density	&	Shape / Nr. of components    \\
\hline
%
Orion\,A	&	206.5	&	-17	&	347 $pc$ \& 393 $pc$	&	Double peak	 \\ 
Orion\,A	&	209		&	-19	&	360 $pc$ \& 387 $pc$	&	Double peak	 \\ 
Orion\,A	&	212.5	&	-19	&	406 $pc$	&	Dominant one peak	 \\ 
Orion\,A	&	214.5	&	-20	&	427 $pc$	&	Dominant one peak	 \\ 
California	&	158		&	-10.5	&	437	&	Dominant peak, additional lower-density peak at 268 $pc$ 	 \\ 
California	&	158.5	&	-12	&	445  &	Dominant peak, additional lower-density peak at 290 $pc$ \\ 
California	&	160.5	&	-9.5	&	455	&	Elongated; 410 - 530 $pc$	 \\ 
California	&	161.5	&	-8.5	&	  -	&	Elongated; 410 - 530 $pc$, no dominant peak	 \\ 
California	&	163		&	-8.5	&	507	&	Dominant peak within the elongated sheet; 410 - 530	 \\ 
California	&	164.5	&	-7.5	&	505	&	Dominant one peak	 \\ 
California	&	165.3	&	-9	&	514	&	Elongated; 455 - 530 $pc$	 \\ 
\hline                              
\end{tabular}
\tablecomments{Our model estimates densities for fixed points in 3D space, therefore there is no direct uncertainty assigned to our distance predictions. However, the predicted densities have full uncertainty estimates that can be propagated into distance estimates \citep[as explained in][]{Rezaei_Kh_20}. Having done so, we estimate uncertainties for our predicted distances to Orion\,A and California clouds to be 10 $pc$.\\
}
\end{table*}

\section*{Acknowledgments}
We would like to sincerely thank the anonymous referee for their thoughtful and constructive comments that improved the quality of the paper. 
SR thanks Henrik Beuther for fruitful discussions. This work has made use of data from the European Space Agency (ESA)
mission {\it Gaia} (\url{https://www.cosmos.esa.int/gaia}), processed by
the {\it Gaia} Data Processing and Analysis Consortium (DPAC,
\url{https://www.cosmos.esa.int/web/gaia/dpac/consortium}). Funding
for the DPAC has been provided by national institutions, in particular
the institutions participating in the {\it Gaia} Multilateral Agreement.
JK has received funding from the European Union's Horizon 2020 research and innovation programme under grant agreement No 639459 (PROMISE).

\begin{appendix}

\begin{figure*}
\resizebox{\hsize}{!}{\includegraphics[clip=true]{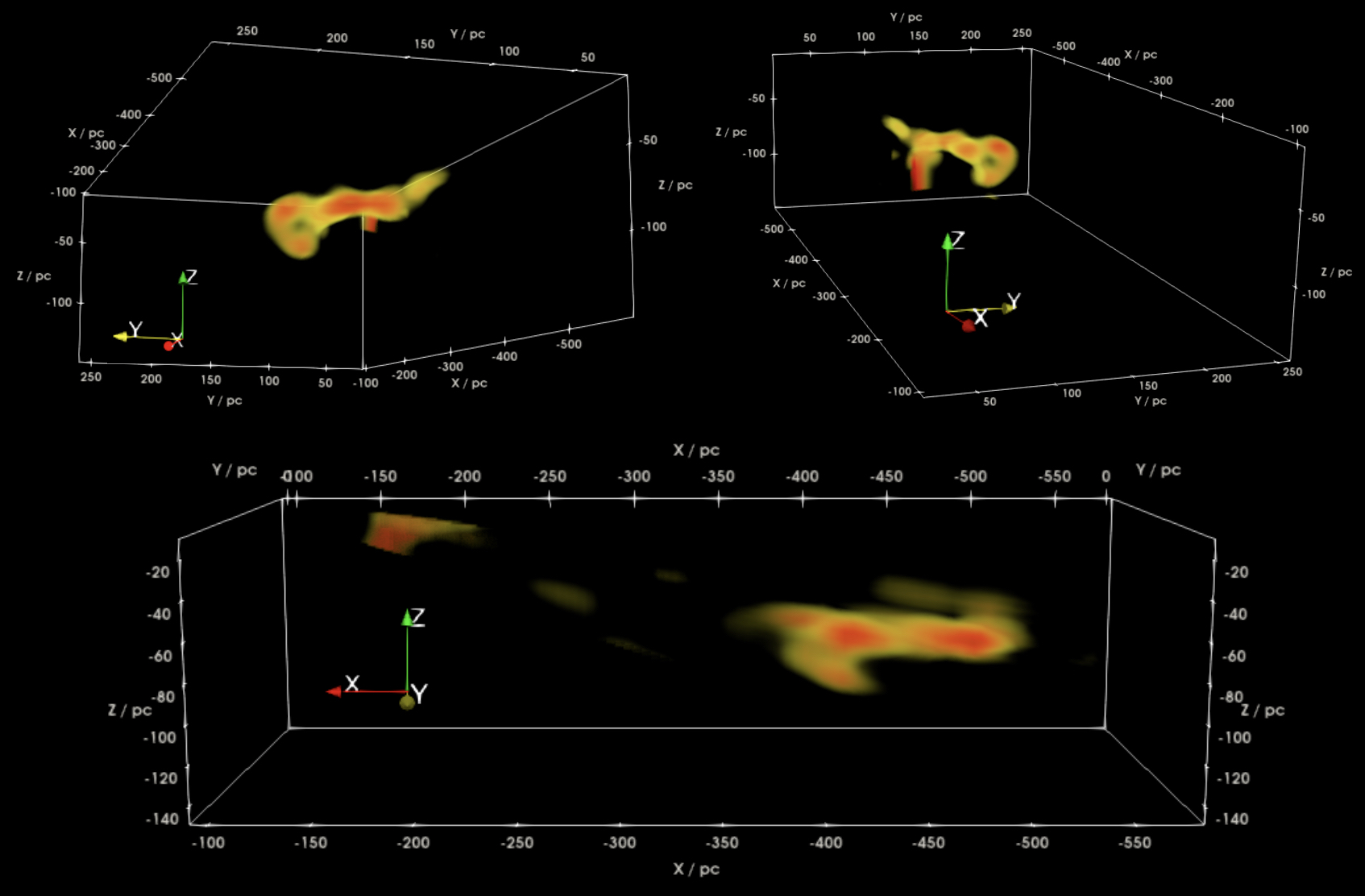}}
\caption{3D structure of California from multiple viewing angles. Small arrows show the orientation of each angle. The bubble described previously and marked in Fig. \ref{fig:3d} is visible in multiple panels. Colour scales are as in Fig. \ref{fig:3d}. \label{fig:3D_more}}
\end{figure*}

\section{Reconstructed extinction}
\begin{figure}
\centering
\begin{minipage}{0.45\textwidth}
\centering
\includegraphics[width=1\textwidth, angle=0]{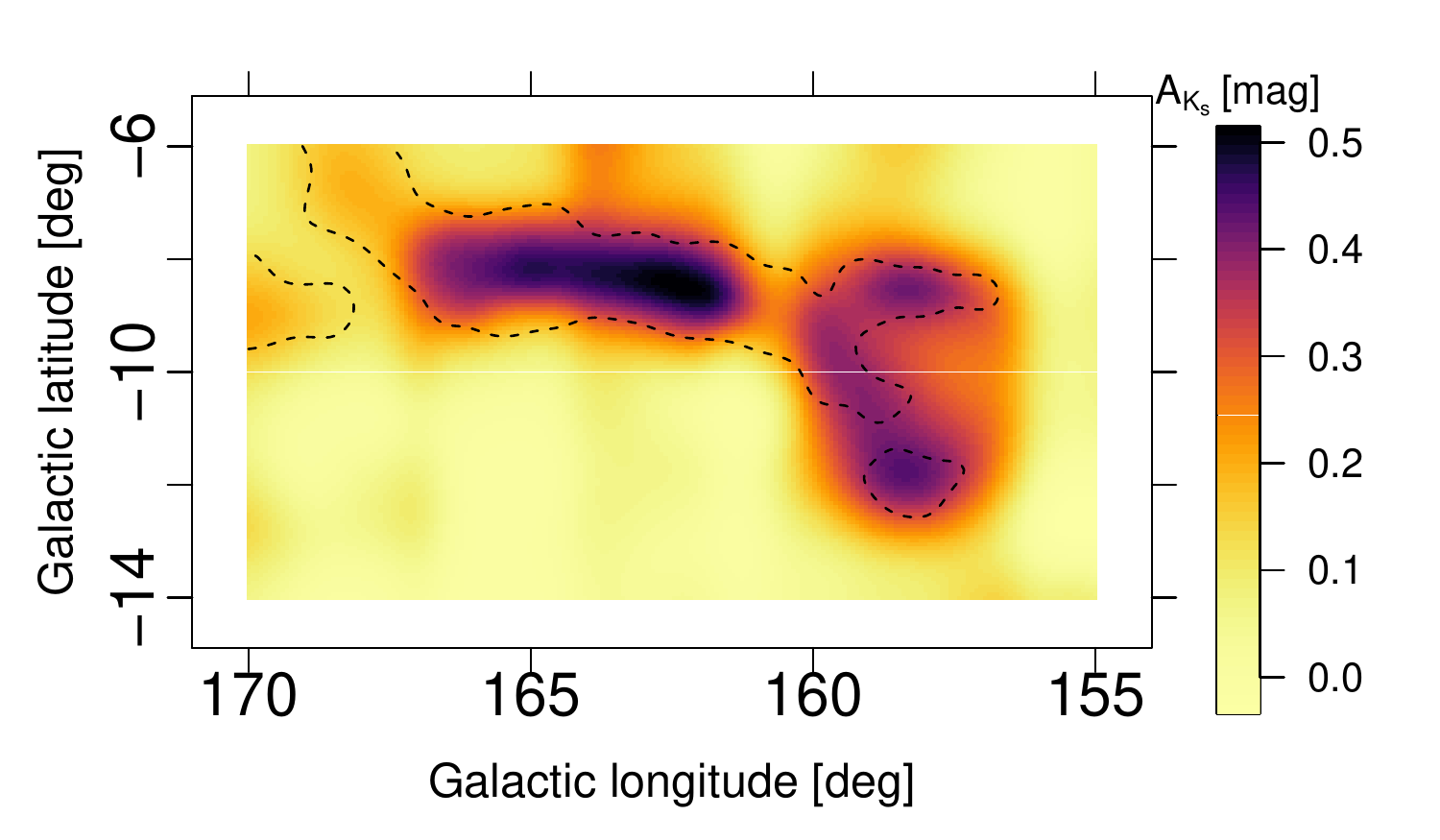}
\caption{Reconstructed extinction using the 3D density predictions (in $K_{s}$ band). Extinctions are calculated using the sum of the predicted 3D densities along each l.o.s. The 2D representation of the 3D predictions nicely recover the plane-of-the-sky features of California. The dashed lines are contours of 0.3 $mag$ extinctions from \cite{Green_19} using ``dustmaps'' interface of \cite{Green_18}. For consistency, the second map is smoothed to the same resolution as our map. \label{fig:Ak}}
\end{minipage}\hfill
\begin{minipage}{0.45\textwidth}
\centering
\includegraphics[width=0.9\textwidth, angle=0]{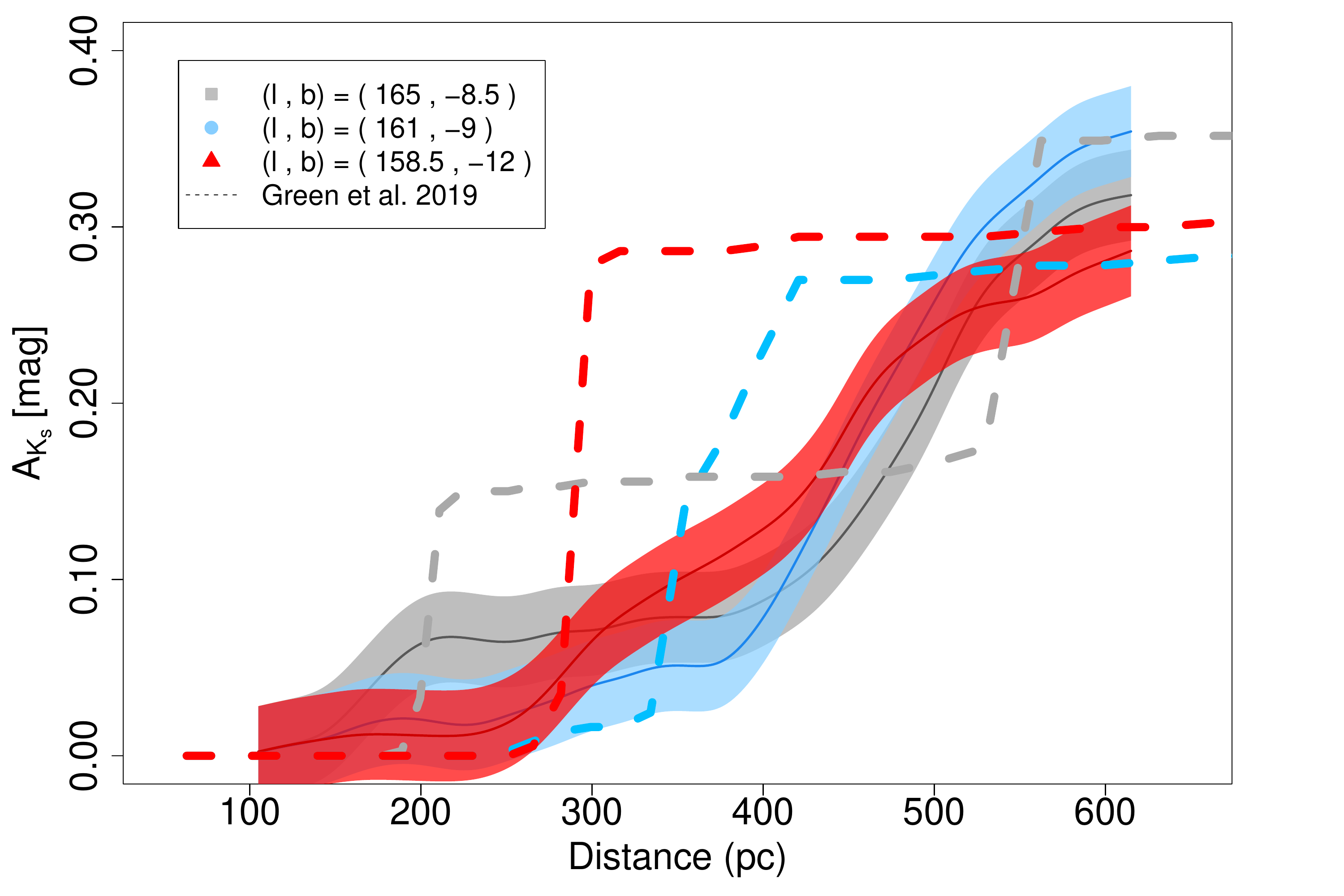}
\caption{Extinction as a function of distance for three l.o.s towards California as shown in figure \ref{fig:los} (solid lines with one sigma uncertainties as shaded area), plus those from \cite{Green_19} (dashed lines). A typical uncertainty in \cite{Green_19} is about 0.02 magnitude in this range. The predicted distances for grey and blue l.o.s agree fairly well in both maps, while the red curve from \cite{Green_19} is missing the second peak in our map. \label{fig:E}}
\end{minipage}
\end{figure}

Figure \ref{fig:Ak} shows the reconstructed extinctions ($A_{K_{s}}$) using the predicted 3D densities. For each l.o.s the predicted densities are summed up, multiplied by the cell sizes in order to get the projected extinctions. Despite the low resolution, the structures resemble that of the higher resolution maps \citep{Lada_09}. Contours of $A_{K_{s}}$ = 0.3 $mag$ from \cite{Green_19} is shown on the figure.

Figure \ref{fig:E} shows the reconstructed extinctions ($A_{K_{s}}$) using the predicted 3D densities as a function of distance for three l.o.s shown in figure \ref{fig:los}, together with those from \cite{Green_19} map. The slope of the reddening curves represents the amount of dust density in the corresponding distance. Therefore, an increase in the redding in figure \ref{fig:E} is equivalent to a peak in our l.o.s density profile in figure \ref{fig:los}. Overall, there is a fairly good agreement between our predicted distances to the clouds and those from \cite{Green_19}; the only difference is for the red curve, where the secondary peak in our map is not recovered by \cite{Green_19}.

Apart from the predicted distances, predicted extinction in \cite{Green_19} seem to appear in steps; allocating almost all predicted dust densities towards a cloud to a certain distance, while an increase in extinction in our map is more gradual. This could be the result of different techniques used in each work: the method used in \cite{Green_19} is not optimal for predicting multiple over-densities near one another or recovering an extended structure. Our technique, on the other hand, tends to smooth the predictions; therefore, underestimating the maximum predicted density for each peak. It is important to note, however, that the amount of elongation caused by the smoothing in our method is less than the 10 $pc$ uncertainties; thus, the predicted elongated structures in our map are real effects of the data rather than an artefact caused by the technique (see section \ref{mock} for an example using mock data).

\section{Mock dataset} \label{mock}
Here we demonstrate the ability of the method to capture the locations and widths of the clouds quite precisely. Also, demonstrate how the peak density predictions in the map is affected by its resolution.

We simulate mock observations using two dust clouds with different sizes and densities, and stars with similar spatial density distribution as that of Gaia. Here are the true properties of the simulated clouds:

- Cloud 1 at $450 pc$, $5 pc$ width, density of $2 {\times} {10}^{-3} mag/pc$

- Cloud 2 at $650 pc$, $70 pc$ width, density of $1 {\times} {10}^{-3} mag/pc$\\
We then calculate the l.o.s extinction to stars in this region and add 20\% noise to both distances and extinctions in order to have a more realistic simulation. We use these as the input data and use the model to find the underlying 3D dust densities.

\begin{figure}
\begin{center}
\includegraphics[width=0.50\textwidth, angle=0]{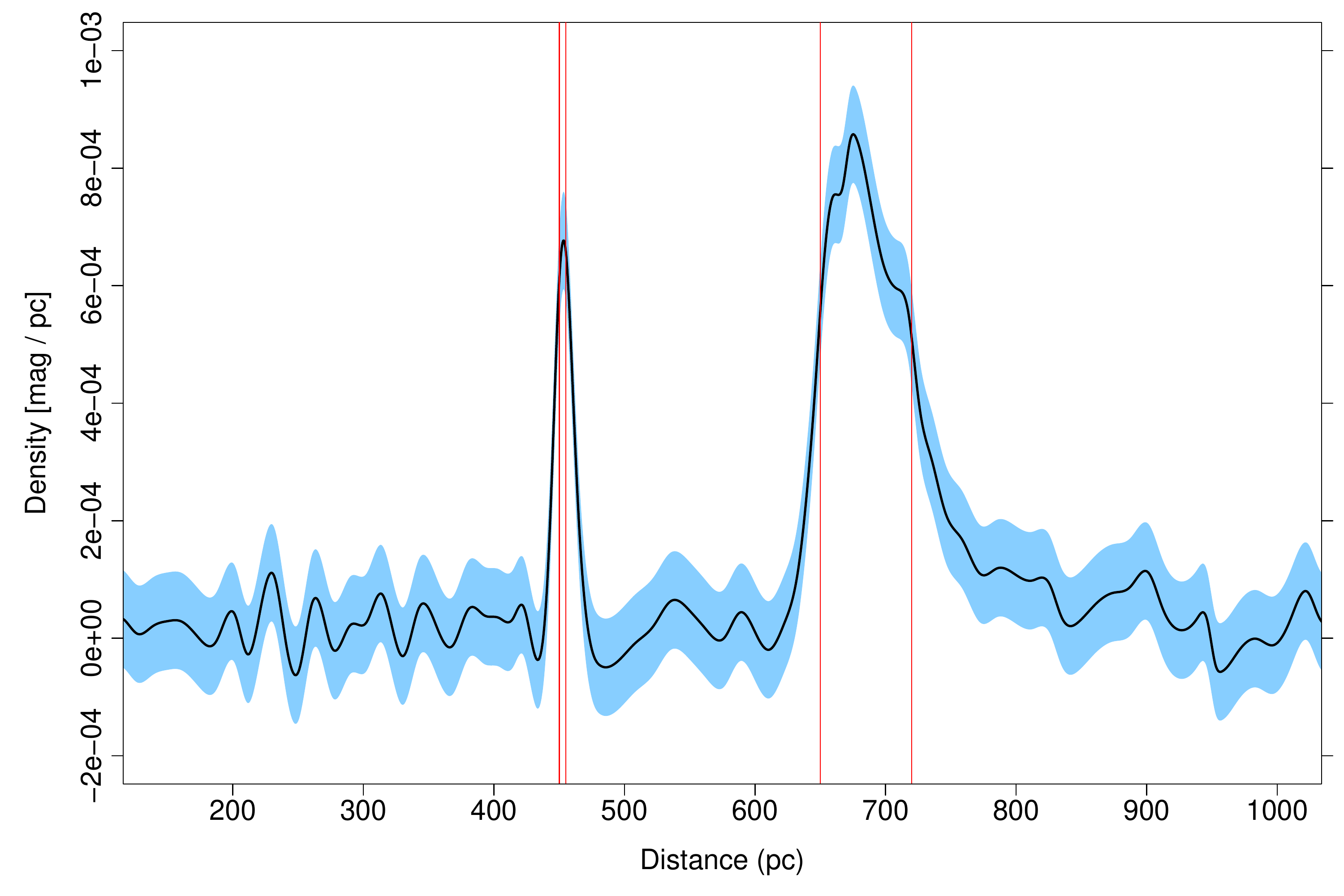}
\caption{Density predictions as a function of distance for mock dataset explained in Appendix \ref{mock}. The vertical red lines demonstrate the ''true`` distances and widths of the clouds. \label{fig:mock}}
\end{center}
\end{figure}

As can be seen in Fig \ref{fig:mock}, the model can predict the location of the clouds and their widths quite precisely. The predicted widths (elongation) of the clouds at the half maximum are 15 $pc$ for the first one and 85 $pc$ for the second one which is consistent with their true widths within $\pm$10 $pc$ uncertainties.
It is worth mentioning that even though the correlation length of the model was 20 $pc$, it recovered clouds with completely different lengths from 5 $pc$ to 70 $pc$; indicating that the elongation of the cloud is mainly determined by the input data. 
Another important point about the predictions are the maximum densities: since the size of the first cloud is smaller than the scale lengths of the model, the maximum predicted density is less than half of the``true" input density. These points indicate the following:
- Our predicted distances are quite precise  
- The extension of the clouds are reliable and determined by the input stars (with the predicted uncertainties)
- The maximum densities probed by the map is an average density within the resolution of the map. The black line is the mean and the blue shaded area the standard deviation of the predictions.

\end{appendix}

\bibliographystyle{aasjournal}
\bibliography{Rezaei_Kh_2022_California_EDR3_final_19_04}

\end{document}